\documentstyle[epsf]{mn}

\def\eq#1 {eq.~(\ref{eq:#1})}

\def\etal{{\it et al.\ }}

\def\eg{{\it e.g.},}
\def\ie{{\it i.e.},}

\def\ltsima{$\; \buildrel < \over \sim \;$}
\def\lsim{\lower.5ex\hbox{\ltsima}}
\def\gtsima{$\; \buildrel > \over \sim \;$}
\def\gsim{\lower.5ex\hbox{\gtsima}}
\def\ga{\mathrel{\hbox{\rlap{\hbox{\lower4pt\hbox{$\sim$}}}\hbox{$>$}}}}
\def\la{\mathrel{\hbox{\rlap{\hbox{\lower4pt\hbox{$\sim$}}}\hbox{$<$}}}}

\def\kms{\,{\rm km\,s{^{-1}}}}
\def\hmpc{\,h{^{-1}}{\rm Mpc}}

\def\la{\langle}
\def\ra{\rangle}

\def\pmb#1{\setbox0=\hbox{#1}%
 \kern-.025em\copy0\kern-\wd0
 \kern.05em\copy0\kern-\wd0
 \kern-.025em\raise.0433em\box0}

\def\vnabla{\pmb{$\nabla$}}
\def\div{\vnabla\!\cdot\!}

\def\divv{\div\vv}

\def\br{{\bf r}}
\def\bv{{\bf v}}

\def\lcdm {$\Lambda$CDM}

\def\hmpc{\, h^{-1} {\rm Mpc}}

\def\3hmpc{\, ( h^{-1} {\rm Mpc})^3}
\def\kms{\, {\rm km\,s^{-1}}}

\def\vv {{ \bf v}}

\def\bs {{ \bf s}}
\def\bd {{ \bf d}}
\def\bo {{ \bf o}}

\def\bR {{ \bf R}}

\def\bH {{ \bf H}}

\def\bepsilon {{ \bf \epsilon}}
\def\bsigma {{ \bf \sigma}}
\def\btepsilon {{ \bf \tilde \epsilon}}
\def\blambda {{ \bf \lambda}}

\def\r0p { r{_0^\prime}}
\def\prior {{\it prior}}

\def\uo{u^o}

\title[Unbiased Reconstruction of the Large Scale Structure]{Unbiased
Reconstruction of the Large Scale Structure}

\author[Saleem Zaroubi]{Saleem Zaroubi\\Max-Planck-Institut f\"ur Astrophysik, 
Karl-Schwarzschild Str. 1, D-85740, Garching, Germany}

\begin{document}

\maketitle
\begin{abstract}
We present a new Unbiased Minimal Variance (UMV) estimator for the
purpose of reconstructing the large--scale structure of the universe
from noisy, sparse and incomplete data. Similar to the Wiener Filter
(WF), the UMV estimator is derived by requiring the linear minimal
variance solution given the data and an assumed \prior\ model
specifying the underlying field covariance matrix. However, unlike the
WF, the minimization is carried out with the added constraint of an
unbiased reconstructed mean field. The new estimator does not
necessitate a noise model to estimate the underlying field; however,
such a model is required for evaluating the errors at each point in
space. The general application of the UMV estimator is to predict the
values of the reconstructed field in un-sampled regions of space
(\eg\ interpolation in the unobserved Zone of Avoidance), and to
dynamically transform from one measured field to another (\eg\
inversion of radial peculiar velocities to over-densities). Here, we
provide two very simple applications of the method. The first, is to
recover a 1D signal from noisy, convolved data with gaps, \eg\ CMB
time-ordered data. The second application is a reconstruction of the
density and 3D peculiar velocity fields from mock SEcat galaxy
peculiar velocity catalogs.

\end{abstract}

\section{Introduction}
\label{intro}

Mapping the distribution of galaxies and their peculiar velocity field
constitutes a major research area in modern astronomy setting both the
observational and theoretical foundations of cosmology and, in
particular, of large scale structure (LSS).  Within the framework of
gravitational instability, the large scale galaxy distribution offers
a probe of the nature of the primordial perturbation field, and can be
used to set strong constraints on the values of cosmological
parameters.  However, since astronomical observations give only
incomplete and noisy information on the real universe, the recovery of
the underlying signal from these observations can be a non-trivial
task, forcing one to resort to regularization methods, \eg, Wiener
filtering (Wiener 1949; Rybicki \& Press 1992; Zaroubi \etal\ 1995) and
the Maximum Entropy algorithm (Gull 1989).

In particular, the WF has been widely applied to galaxy surveys (Lahav
\etal\ 1994; Fisher \etal\ 1995; Schmoldt \etal\ 2000), CMB studies
(Bunn \etal\ 1994; Tegmark \& Efstathiou 1997, Bouchet \& Gispert,
1999), and galaxy peculiar velocity catalogs (Hoffman \etal\ 2000;
Zaroubi 2000; Zaroubi \etal\ 1999 \& 2001a). The WF provides an
optimal estimator of the underlying field in the sense of a
minimum-variance solution given the data and an assumed \prior\ model
(Wiener 1949; Press \etal\ 1992). The \prior\ defines the data
auto-correlation and the data-field cross-correlation matrices. In the
case where the data is drawn from a random Gaussian field, the WF
estimator coincides with the conditional mean field and with the most
probable configuration given the data (see Zaroubi \etal\ 1995).

Although the application of the WF is very simple and has proven to be
useful for many purposes, it is easy to show that the estimator is
intrinsically biased, often in a scale dependent manner.  The main
cause of this bias stems from the modulation introduced by the
Signal/(Signal + Noise) weighting it invokes. This drawback has
prevented the use of the Wiener reconstructed maps in many areas, \eg\
power spectrum estimations, bias parameter extraction from galaxy
peculiar velocity data comparison with Galaxy surveys data, etc.  To
account for this bias in the power spectrum estimation a correction
factor is often applied to the Wiener reconstructed signal (Rybicki \&
Press 1992; Tegmark \& Efstathiou 1997).

In this paper, we propose a new linear unbiased minimal variance (UMV)
estimator that is designed to avoid the intrinsic bias that exists in
the WF. This is achieved by solving the minimization equation subject
to the constraint of unbiased mean underlying field. To test the UMV
estimator we apply it: 1- To recover a 1D time series from convolved
and noisy data with gaps. 2- To mock SEcat peculiar velocity data set,
a combination of the SFI (Giovanelli \etal\ 1998) and the ENEAR (da
Costa \etal\ 2000) galaxy peculiar velocity catalogs, where we
reconstruct the distribution of the true density and velocity fields
from which the mock catalog is constructed.

The outline of this paper is as follows.  The method is presented in
\S~\ref{sec:theory}.  In \S~\ref{sec:regularization} two simple extra
regularization methods are discussed. \S~\ref{sec:CR} shows how
constrained realizations could be produced within the UMV framework.
The method is tested using artificial data based on simulations in
\S~\ref{sec:applications}. The results are discussed and the
conclusions are summarized in \S~\ref{sec:summary}.

\section{Theory}
\label{sec:theory}

\subsection{Derivation of the UMV estimator}
\label{sec:umv}

Consider the case of a set of observations, or measurements, performed
on an underlying field ${\bf s}=\{s_\alpha\} (\alpha=1,...,N)$, or on
any field linearly related to {\bf s}, which yields a set of data
points, ${\bf d}=\{d_i\} (i=1,...,M)$. In particular, we are interested in
measurements that can be modeled mathematically as a linear
convolution or mapping of the underlying field,

\begin{equation} 
{\bf d} =\bo +\bepsilon = {\bf Rs} +\bepsilon\;\; ,
\label{eq:basic}
\end{equation} 
where ${\bo}= {\bf Rs}$ and ${\bf R}$ is an $M\times N$ matrix which
represents the response or point spread function (hereafter RF) and
$\bepsilon$ is the noise vector associated with the data. The RF will
be treated as any function that linearly connects the underlying
signal to the data, be it blurring or smoothing introduced by the
measurement, some theoretical relationship between two fields, or any
other linear transformation on the underlying signal.

In principle, reconstructing ${\bf s}$ can be accomplished by
inverting equation eq.~\ref{eq:basic}. However two main obstacles
usually prevent one from pursuing this approach. First, the number of
independent data points is usually much smaller than the number of
underlying degrees of freedom. Second, the presence of noise can
render such a direct inversion unstable and the obtained results
meaningless.  Due to these potential difficulties one is often forced
to resort to some statistical regularization techniques (\eg\ WF) in
order to solve eq.~\ref{eq:basic}.

Similar to the requirements of the derivation of the WF we assume the
\prior\ knowledge of the signal correlation function
\begin{equation} 
{\bf S}=\langle \bs\, \bs^+\rangle,
\label{eq:corr} 
\end{equation} 
where $\bs^+$ is the complex conjugate of the transpose of the
underlying signal, and angled brackets denote the signal ensemble
average. Notice, that there have been no assumptions made regarding
the actual probability distribution function from which the field
$\bs$ is drawn. We define the unbiased minimal variance estimator,
$\bs^{UMV}= {\bf H} \bd$ where ${\bf H}$ is an $N\times M$ matrix that
minimizes the variance of the residual ${\bf r} = \bs - \bs^{UMV}$,
while satisfying the constraint,
\begin{equation}
\lbrack \bs^{UMV} \rbrack_{{\cal N}}=\lbrack {\bf H}
\bd\rbrack_{{\cal N}} = \bs.
\label{eq:constraint}
\end{equation}
The ensemble average, $\lbrack \dots \rbrack_{{\cal N}}$, is an ensemble
average over noise realizations and is very different from the
ensemble average, $\langle \dots \rangle$, which denotes an ensemble
average over signal realizations.  This distinction between the two
ensemble averages will be used only in this section, in the rest of
the paper we use only the signal ensemble average.

In short, we are seeking ${\bf H}$ that minimizes,
\begin{equation} 
\langle\lbrack{\bf r r^+} + \blambda \bH \bd
\rbrack_{{\cal N}}\rangle=\langle\lbrack (\bs - \bs^{UMV}) (\bs - \bs^{UMV})^+ +
\blambda \bH \bd\rbrack_{{\cal N}}\rangle
\label{eq:minimize}
\end{equation} 
where $\blambda$ is a {\it Lagrange multiplier}. The ensemble average
over noise realizations precedes the one over signal realization. The
order of averaging over ensembles is very important since if it was
reversed the contribution of the term $\bH \bd$ will be null.

The constraint, introduced in eq.~\ref{eq:constraint}, assumes that
the data is unbiased, namely the errors are random, and therefore
requires that the estimator does not alter the value of the measured
data points but rather it forces the field to retain the measured
values at their appropriate locations. However, this requirement does
not guarantee an unbiased variance of the reconstruction, on the
contrary one expects that the variance of the reconstructed field is
some compromise between the (Signal + Noise) variance of the data
points and the assumed variance of the underlying signal.

Carrying out the minimization of eq.~\ref{eq:minimize} with respect to
\bH\, one obtains an equation that together with eq.~\ref{eq:constraint}
are used to solve for $\bH$ and $\blambda$. The solution yields the
UMV estimator,
\begin{equation}
\bs^{UMV} = \langle \bs \bo^+ \rangle \langle\bo \bo^+\rangle^{-1} \bd.
\label{eq:estimator}
\end{equation} 
The {\it Lagrange multiplier}, $\blambda$, is roughly proportional to
the {\it Noise/(Signal+Noise)} making it dominant when the noise is dominant
and small when the Signal/Noise $\gg$ 1.

In the absence of a response function that operates on the signal,
namely, ${\bf R=I}$ (where {\bf I} is the unity matrix), the
reconstructed signal at the location of the data points is identical
to the data measured values which is consistent with the constraint
given in Eq.~\ref{eq:constraint}.  In the rest of space the degrees of
freedom are recovered by interpolating the data points in a manner
consistent with the correlation assumed in the underlying theory.

Mathematically, the difference between the WF and the current
estimator is that in the former the term $\langle\bo \bo^+\rangle$ is
replaced by $\langle\bd \bd^+\rangle$, a matrix that includes the
noise correlations, an addition that accounts for the signal
suppression which renders the WF mean field biased.

The variance of the field estimated in Eq.~\ref{eq:estimator} is,
\begin{eqnarray} 
& \la\bs^{UMV} & {\bs^{UMV}}^+ \ra  = \nonumber \\ & & \la \bs
\bo^+\ra \la \bo \bo^+\ra^{-1}\left(\la \bo \bo^+\ra + {\bf N}\right)
\la \bo^+ \bo\ra^{-1} \la \bo \bs^+\ra\quad,
\label{eq:variance}
\end{eqnarray}
where {\bf N} is the noise correlation matrix. 
The variance of the residual,
\begin{eqnarray}
\la {\bf r r^+} \ra & = & \la \bs \bs^+\ra - \la \bs \bo^+\ra \la \bo
\bo^+\ra^{-1} \la \bo \bs^+\ra \nonumber\\ & & + \; \la \bs \bo^+\ra
\la \bo \bo^+\ra^{-1} {\bf N} \la \bo \bo^+\ra^{-1} \la \bo \bs^+\ra
\quad.
\label{eq:resvar}
\end{eqnarray} 

As a simple example, assume that ${\bf R=I}$ and that the field is
estimated in the exact locations of the data points, the variance in
eq.~\ref{eq:variance} simply reduces to $\la \bs \bs^+\ra + {\bf N}$,
recovering the power spectrum of the signal + noise at those
points. The variance of the residual (eq.~\ref{eq:resvar}) at the
location of the data points is simply reduced to the correlation
matrix of the noise, {\bf N}. In addition, when the data points are
uncorrelated with the rest of the underlying degrees of freedom then
the reconstructed values, at locations different from those of the
data points, are zero (eq.~\ref{eq:variance}) and the variance of the
residual at those same locations, as obtained from
eq.~\ref{eq:resvar}, is simply the underlying \prior\ correlation.

Substitution of $\bo= \bR\bs$ in
equation~\ref{eq:estimator} yields,
\begin{equation}
\bs^{UMV} = \langle \bs \bs^+{\bf R}^+ \rangle \langle{\bf Rs} \bs^+{\bf
       R}^+\rangle^{-1} \bd,
\end{equation}
which one could be tempted to further simplify to obtain $\bs^{UMV} =
\langle \bs \bs^+\rangle \langle \bs \bs^+ \rangle^{-1}{\bf R}^{-1}
\bd = {\bf R}^{-1} \bd$.  Normally however, the matrix $\bR$ is not
square but rather has a larger number of rows than columns and its
inverse is not unique and the inverse of its transpose, $\bR^+$, does
not exist at all. Therefore, in most cases carrying the simplification
further is mathematically incorrect. This simplification is possible
if the inverse of both $\bR$ and $\bR^+$ exist, which is only true if
the number of data points is identical to the number of degrees of
freedom in the underlying signal or when the signal correlation
function is a Dirac-delta function. 

However, when the simplification could be performed the UMV estimator
is equivalent to direct inversion.  Since direct inversion in the
presence of noise could produce very large uncertainties in the
reconstructed signal one should resort to extra regularization that
reduces the variance to a manageable size, an example of such an extra
regularization is given in \S~\ref{sec:regularization}.

To summarize, the regularization strength of the UMV estimator stems
from the cross talk between the underlying signal and the data points
at different locations. The stronger the correlation is, the more
advantageous the use of the UMV. With the absence of such a cross talk
the number of degrees of freedom that one can reconstruct is identical
to the number of data points, and the UMV estimator is reduced in this
case back to, $\bR^{-1}$, namely equivalent to direct inversion, a
case that requires extra regularization in order to reduce the,
often huge, uncertainty introduced by the noise.

\section{Extra Regularization}
\label{sec:regularization}

As shown in the previous section, the UMV could be equivalent to
direct inversion, therefore, while it produces an unbiased estimator
of the underlying field, it gives a ``minimum variance'' that is too
large to be useful. In contrast, the WF filter produces a manageable
variance but a large degree of bias. Hence, the UMV alone is often
insufficient to stabilize the deconvolution and consequently it must
be either modified, supplemented with extra regularization, or
replaced with another method.

Of course, to fully solve the inversion problem one might want to use
more sophisticated non-linear methods such as Maximum Entropy (Gull
1989), Pixons image restoration algorithm (Pi\~{n}a \& Puetter 1993),
etc. These methods, however more advanced, are much more
computationally expensive and complicated to apply.

In this paper we choose to modify the UMV inversion method with extra
regularization that is both simple and easy to integrate in the
algorithm.  In subsections~\ref{sec:svd-regularization}
\&~\ref{sec:wf-regularization}, two simple extra regularization
methods are introduced, the Singular Value Decomposition (SVD)
algorithm (\eg\ Press \etal\ 1992), and a WF-like
regularization. These methods are basically applied to find the vector
${\bf a}$, defined as ${\bf a} \equiv {\bf O}^{-1}
{\bd}\equiv\langle\bo \bo^+\rangle^{-1} \bd $ (see
Eq.~\ref{eq:estimator}). The difficulty in finding {\bf a} could be
due to the inversion instability of {\bf O} or due to the noise in
matrix {\bf a}. In both methods suggested here this difficulty is
solved by, in effect, adding elements to the diagonal of {\bf O} or to
its eigen-values so that it stabilizes the inversion of {\bf O} and at
the same time suppress the noise contribution of {\bf d} to the vector
{\bf a}.

\subsection{The Singular Value Decomposition Algorithm as a Regularizer}
\label{sec:svd-regularization}

The following presentation follows the one in Zaroubi \etal\ (1995).
Mathematically the solution of Eq.~\ref{eq:estimator} involves the
solution of the equation
\begin{equation}
 {\bf O a} = {\bd},
\label{eq:inverse}
\end{equation}
for the unknown vector ${\bf a}$. 
matrix.

The SVD algorithm basically decomposes the positive definite matrix
${\bf O}$$^\ast$\footnote{The matrix ${\bf O}$ is an
auto-correlation matrix therefore it is a square positive definite
matrix.} into a multiplication of three matrices, ${\bf O = U}\ {\rm
diag} \lbrace w_i \rbrace\ {\bf U^+}$, where the set $\lbrace w_i\rbrace
$ is referred to as the collection of singular values, or the
eigenmodes, and the columns of the matrix ${\bf U}$ are orthogonal and
proportional to the eigenvectors of ${\bf O}$. The inversion, after
decomposition, is straightforward and gives ${\bf O}^{-1} ={\bf U}\
{\rm diag} \lbrace 1/w_i\rbrace\ {\bf U^+}$. Formally speaking,
Eq.~\ref{eq:inverse} has a unique solution if and only if ${\bf O}$ is
a non-singular matrix, namely if $w_i\neq 0$ for all $i$.  However a
meaningful solution can be obtained even in the case where ${\bf O}$
is singular, by requiring the solution to minimize the norm
of the residuals, $|{\bf O a} -\bd|$. Such a solution is obtained by
substituting $1/w_i=0$ in the expression for the inverse for any
$w_i=0$ (Press \etal\ 1992).

The question arises in a particular problem of setting the lower limit
of the singular values, below which the inverse values are set to
zero.  In general, the singular values measure the amount of
`information' carried by each mode in the problem (Press \etal\ 1992),
namely the small singular values do not have significant contribution
to reconstruction, nevertheless, they can destabilize the
inversion. As an extension of the ideal case of $w_i=0$, we impose a
cutoff on the small singular values in order to maintain
stability. Often, the structure of the sorted spectrum of singular
values contain a very sharp drop normally appearing as a 'knee'. The
location of the 'knee' usually determine the singular values that
contain sufficient information. A simple cutoff at the location of the
'knee' normally does the trick and stabilizes the inversion.  

Another possible criterion for the choice of the cutoff would be by
expanding the matrix {\bf O} in terms of signal-to-noise eigenmodes
(Bond 1995).  This method allows a simultaneous diagonalization of the
Matrix {\bf O} and the noise matrix {\bf N} which in turn allows the
usage of some signal-to-noise ratio threshold (typically 1) in order
to set the cutoff. For specific application see the 1D example in
\S~\ref{sec:1d_example}, especially the expansion shown in
Fig.~\ref{fig:svd} (see also the the example discussed in Zaroubi
\etal\, 1995).

\subsection{Wiener Filter-like Regularizer} 
\label{sec:wf-regularization}

The WF by itself is a very robust and efficient regularizer. The main
reason for this stems from the relatively high values of the diagonal
terms in the matrix $\langle\bd \bd^+\rangle$. These high values
usually come from the noise contribution, especially if it is
uncorrelated (\eg\ white noise). This aspect is, of course, also
responsible for the suppression of the Wiener reconstructed signal.

From the point of view of the SVD algorithm the stabilization effect
caused by the noise diagonal elements is quite obvious, especially in the
light of the following example. Consider the simplest case of white
Gaussian noise where the noise correlation matrix is proportional to
the unity matrix, ${\bf I}$. With the application of the SVD algorithm
the noise contribution is still proportional to the unit matrix and is
the same for every singular value, this means that the singular values
can be as small as the noise and non of them has a value of 'zero',
therefore, the matrix is naturally stable an no extra stabilization is
normally needed.

In order to utilize the stabilization aspect of the WF on the one
hand, while avoid the signal suppression aspect on the other, we add a
very small white noise-like contribution to the correlation matrix
${\bf O}$. Naturally, one should seek the smallest possible addition
in order to suppress the recovered signal as little as possible. The
choice of the amplitude of the addional noise term could be guided by
the same signal-to-noise cutoff criterion discussed in the previous
section (\S~ref{sec:svd-regularization}). This is, of course, problem
dependent and should be carried out with caution.

An example of the application of this approach is discussed in
\S~\ref{sec:1d_example}.

\section{Constrained Realizations}
\label{sec:CR}

Hoffman \& Ribak (1991) showed that, within the framework of Gaussian
random fields and a given \prior, any realization can be split into
two parts: the mean field, or the WF field, which is determined by the
constraints (data) and the residual field which is a Gaussian random
field. Thus, one can make a constrained realization of the underlying
field given the assumed {\it prior} model and the data.

With the UMV reconstructed field one can produce constrained noisy
realizations at every point in space only if the noise in the observed
quantity is uncorrelated. The formulation here is quite similar to the
constrained realizations procedure of Hoffman \& Ribak (1991); one
produces a random realization of the noisy underlying field and noise,
${\bf \tilde s^{Noisy}} \left( = {\bf \tilde s}+{\bf\tilde
\sigma}\right) $, then sample it in the same way the actual data is
obtained,

\begin{equation} 
 {\bf \tilde d}= \bR {\bf \tilde s} +  \btepsilon .
\label{eq:randomdata}
\end{equation} 

Here, ${\bf\tilde s}$ and $\btepsilon$ are random realizations of the
underlying field and the statistical uncertainties, respectively. The
noise in the random realization of the data ${\bf \tilde \epsilon}$ is
related to the noise in the random realization of the underlying noisy
field ${\bf\tilde \sigma}$ through the response function,
\begin{equation}
 {\bf\tilde \epsilon} = {\bf R \tilde \sigma}.
\end{equation}

A constrained realization of the field given the data is given by:

\begin{equation}
{\bf s^{Noisy}}=  {\bf \tilde s^{Noisy}} + {\bf H} ({\bf d} -
{\bf \tilde d}).
\label{eq:CR}
\end{equation}
Here ${\bf s^{Noisy}}$ is the constrained noisy realization. 

Note that if the noise, $\btepsilon$, is correlated one cannot use this
formalism as the UMV takes into account only the correlation in the
underlying signal. In such a case, one can think of a different approach,
where the noise correlation is included in the estimator, in this case
one can expand the UMV estimator to have the form,
\begin{equation}
\bs^{UMV} = \langle \bs \bo^+ + \bsigma \bepsilon^+ \rangle \langle\bo
\bo^+ + \bepsilon \bepsilon^+\rangle^{-1} \bd.
\label{eq:noisyestimator}
\end{equation} 

This kind of noisy constrained realization could be very useful for
cases when one has gaps in the data that one would like to fill, not
only in a manner consistent with the data but also with the same power
spectrum of the Signal+Noise.

\section{Applications}
\label{sec:applications}

The UMV estimator can be applied to many areas of LSS and CMB for: 1)
interpolating between data points, \eg\ filling in gaps in the
time-ordered data to help map-making, bridging the Zone of Avoidance
and other uncovered areas in nearly full sky catalogs, etc.; 2)
stabilize inversions used to transform one dynamical field to another,
\eg\ from radial peculiar velocity field to density field; or from
redshift to real density.

Here we present two examples, the first is a reconstruction from one
dimensional time series of noisy convolved signal with gaps, \eg\
CMB time-ordered data. The second example shows how this method could
be applied to reconstruct the 3D density and peculiar velocity fields
from observed galaxy radial peculiar velocity catalogs.

\subsection{Time Series Example: Deconvolution of Noisy Data With Gaps}
\label{sec:1d_example}

\begin{figure}
\setlength{\unitlength}{1cm} \centering
\begin{picture}(8,7)
\put(9.2, -1){\includegraphics{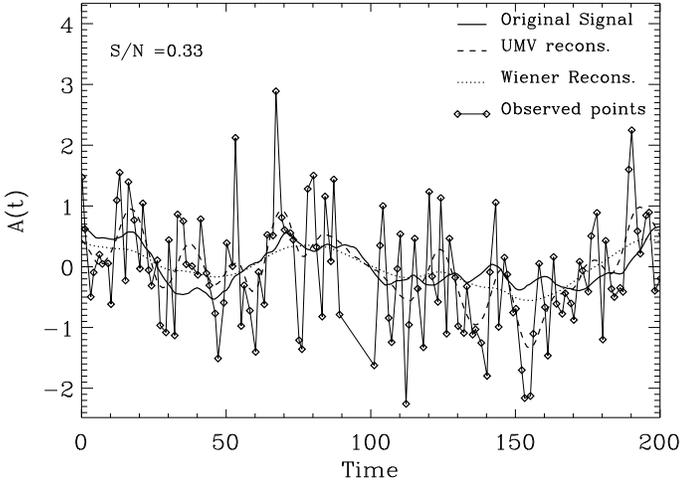}}
\end{picture}
\caption{A one dimensional reconstruction example. The heavy-solid
line shows the underlying signal $A(t)$ as a function of time; both
time and amplitude has arbitrary units. The underlying signal is
convolved with a Gaussian window function with width of 5 time units;
the convolved signal is then uniformly sampled with the exception of a
gap in the time range of 90-100. A random noise was then added to
produce the 'data' points shown with the diamond-shaped connected
points; the signal-to-noise ratio in this example is 3.  The
heavy-dashed line shows the UMV reconstructed signal, while the
heavy-dotted line shows the Wiener reconstructed signal}
\label{fig:1d_conv}
\end{figure}

To test the performance of the UMV estimator we proceed with a simple
one dimensional example. Let ${\bf s}$ be a random Gaussian time
series, with a known correlation function, which we would like to
measure in the time range $[0-200]$ (the signal and time units are
arbitrary). The measurement involves a convolution of the signal with
Gaussian window of 5 time units width, the convolution is
describe by the matrix ${\bf C}$. The measurement procedure uniformly
samples the signal at about 100 positions, except for the time range
of $[90-100]$ where there is a gap in the data. An instrumental white
noise, $\bepsilon$, with standard deviation three times larger than
the signal standard deviation is added. Mathematically the data is
connected to the underlying signal with $\bd = {\bf C} \bs +
\bepsilon$. The heavy-solid line in Fig.~\ref{fig:1d_conv} shows the
underlying signal and the connected diamonds represent the measured
data.

This specific example is typical to what one obtains in CMB
time-ordered data type of measurements where the matrix ${\bf C}$ is
the instrument's point spread function, the noise level reflects the
detectors sensitivity and the gaps are drop-outs in the data stream.

The UMV and WF estimators for this case are,
\begin{equation}
\bs^{UMV} = \langle \bs \bs^+{\bf C}^+ \rangle \langle{\bf Cs} \bs^+{\bf
       C}^+\rangle^{-1} \bd,
\label{eq:1d_conv_umv}
\end{equation}
and,
\begin{equation}
\bs^{WF}  = \langle \bs \bs^+{\bf C}^+ \rangle \langle{\bf Cs} \bs^+{\bf
       C}^+ + \epsilon^2 {\bf I}\rangle^{-1} \bd. 
\label{eq:1d_conv_wf}
\end{equation}

\begin{figure}
\setlength{\unitlength}{1cm} \centering
\begin{picture}(8,8.3)
\put(-1.8, -8){\includegraphics{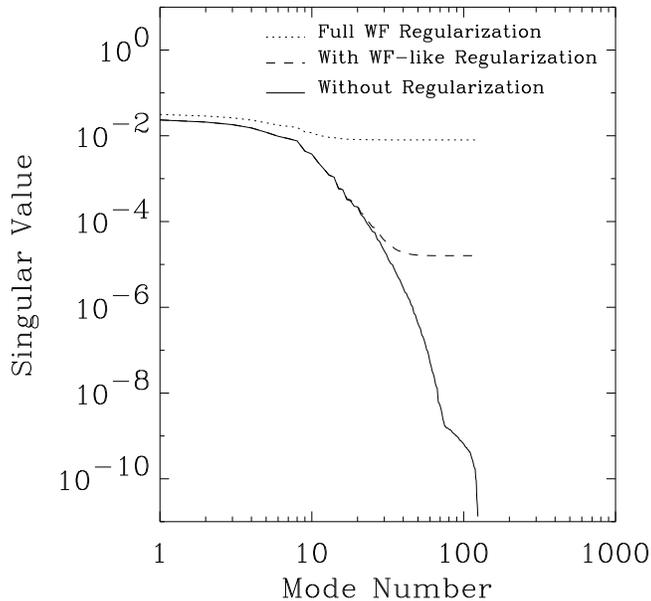}}
\end{picture}
\caption{Sorted spectrum of singular values of the matrix ${\bf
O}$. The heavy-solid line shows the singular values of ${\bf O}$
without regularization. The heavy-dashed line shows the singular
values of ${\bf O}$ after adding a diagonal constant noise matrix, the
constant is $0.005$ of the noise rms. The `knee' around singular values
of $10^{-9}$ is caused by numerical noise. For comparison, the dotted
line shows the singular values of the full WF regularization}
\label{fig:svd}
\end{figure}
Due to the large correlation length in this example the matrix ${\bf
O} \equiv \langle{\bf Cs} \bs^+{\bf C}^+\rangle$ in
equation~\ref{eq:1d_conv_umv} is unstable for inversion, therefore as
previously discussed, one has to apply an extra regularization
scheme. The heavy-solid line in Figure~\ref{fig:svd} shows the sorted
spectrum of the singular values $(w_i)$ of the matrix ${\bf O}$ versus
the mode number. The very abrupt drop of the singular values around
mode number 10 indicates that the information content of the rest of
the singular values is very small and that they are essentially
responsible for the inversion instability. To stabilize the inversion,
we adopt the regularization method described in
\S~\ref{sec:wf-regularization} and add a diagonal constant with
$0.005$ of the noise contribution. The heavy-dashed line in
Figure~\ref{fig:svd} shows the sorted spectrum of singular values of
the regularized ${\bf O}$. The two lines are identical for the large
singular values and depart at small singular values. For comparison,
the dotted line in Figure~\ref{fig:svd} shows the singular values of
the matrix used in the WF application, namely, ${\bf O + N}$.

The heavy-dashed line in Fig.~\ref{fig:1d_conv} shows the UMV
reconstruction of the signal that the follows directly from
calculating $\bs^{UMV}$ from equation~\ref{eq:1d_conv_umv}.  The
dotted line shows the WF reconstructed signal as obtained from
equation~\ref{eq:1d_conv_wf}.  The UMV reconstruction follows the data
points while taking into account the deconvolution and the {\it prior}
for the signal temporal correlations. The WF reconstruction is much
smoother and has smaller variance than the underlying signal.

\begin{figure}
\setlength{\unitlength}{1cm} \centering
\begin{picture}(8,7)
\put(9.2, -1.){\includegraphics{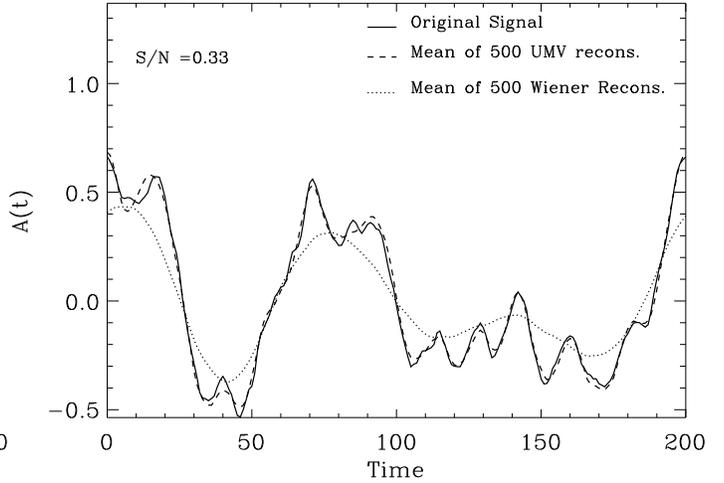}}
\end{picture}
\caption{The solid line shows the same underlying signal as the in
fig.~\ref{fig:1d_conv}. To this signal we add 500 noise realizations
to produce Monte-Calors of the 'observed' data. The dashed line shows
the mean of the 500 UMV reconstructions of these realization. The
dotted line shows the mean of the Wiener reconstruction of the each of
each of the 500 data realizations}
\label{fig:1d_conv_mean}
\end{figure}

To demonstrate the differences between the two reconstructions,
fig.~\ref{fig:1d_conv_mean} shows an average of 500 reconstructions of
data realizations with the same underlying signal but different noise
Monte-Carlos, the unbiased nature of the UMV and the biased nature of
the WF reconstructions are evident.  In each UMV reconstruction we
have applied the aforementioned WF-like regularization, the amount of
bias introduced by this procedure is negligible, while the bias for the
full WF application is not.

\subsection{Reconstruction from Peculiar Radial
Velocity Data}
\label{sec:Radial Velocity}

Next, we present an example of an application of the UMV estimator to
reconstruct the density and 3D velocity fields from galaxy radial
peculiar velocity data. The data set used here is a mock catalog that
mimics the SEcat galaxy peculiar velocity catalog (Zaroubi 2000),
which is a combination of the grouped and Malmquist biases corrected
SFI (Giovanelli \etal\ 1998) and ENEAR (da Costa \etal\ 2000) galaxy
peculiar velocity catalogs.  The catalog consists of about 2050
objects ($\approx 1300$ from SFI and $\approx 750$ from ENEAR) for
which it provides radial velocities and inferred distances with
errors, on the order of $\approx 19 \%$ of the distance per
galaxy. The sampling is reasonably homogeneous and covers the whole
sky outside the ZoA, the radial selection function is uniform out to
$\approx 5500 \kms$. The ZoA is about $\pm 15^\circ$. The SEcat
catalog contains distances and peculiar velocities of both late-type and
early-type galaxies, and therefore has the advantage of sampling both
high and low density environments minimizing possible biases that may
affect reconstruction from catalogs based on a single population of
galaxies.  

The mock catalog is produced as follows, we first generate a random
linear Gaussian realization of density and velocity with a \lcdm\
power spectrum (with $\Omega_m=0.3 \;\&\; \Lambda=0.7$). Then the mock catalog
is produced so that the locations of the data points in the mock
catalog are the same as for those in the real SEcat catalog, however,
the values of the radial peculiar velocities are taken from a random
Gaussian realization of the underlying fields. The original density
field is used for comparison with the reconstructed one.

In this example the data points are given as a set of observed radial
peculiar velocities $\uo_i$ sampled at positions $\br_i$ with
estimated errors $\epsilon_i$, assumed to be uncorrelated. The
observed velocities are thus related to the true underlying velocity
field $\bv(\br)$, or its radial component $u_i$ at $\br_i$, via

\begin{equation} 
\uo_i =\bv(\br_i) \cdot \hat \br_i + \epsilon_i \equiv u_i + \epsilon_i .
\label{eq:eps}
\end{equation}

We assume that the peculiar velocity field $\bv(\br)$
and the density fluctuation
field $\delta(\br)$ are related via linear gravitational-instability theory,
$\delta = f(\Omega)^{-1} \divv$, where $f(\Omega)\approx\Omega^{0.6}$
and $\Omega$ is the mean universal density parameter.
Under the assumption of a specific theoretical prior for the power
spectrum $P(k)$ of the underlying density field,
we can write the UMV estimator of the fields as
\begin{equation}
\bv^{UMV}(\br) = \Bigl < \bv(\br) u_i \Bigr >  \Bigl < u_i u_j \Bigr >
^{-1}        \uo_j,
\label{eq:UMVv}
\end{equation}
and,
\begin{equation}
\delta^{UMV}(\br) = \Bigl < \delta(\br) u_i \Bigr >  \Bigl < u_i u_j
\Bigr > ^{-1}    \uo_j .
\label{eq:UMVd}
\end{equation}
Assuming linear theory and that the velocities are drawn from a
Gaussian random field, the two-point velocity-velocity and
density-velocity correlation tensors (bracketed quantities in
eqs.~\ref{eq:UMVv} \&. ~\ref{eq:UMVd}) are readily calculated.
The calculation of these matrices is discussed elsewhere 
(G\'orski 1988; Zaroubi \etal\ 1995,1999).

\begin{figure}
\setlength{\unitlength}{1cm} \centering
\begin{picture}(8,9.3)
\put(-2.3, -2.5){\includegraphics{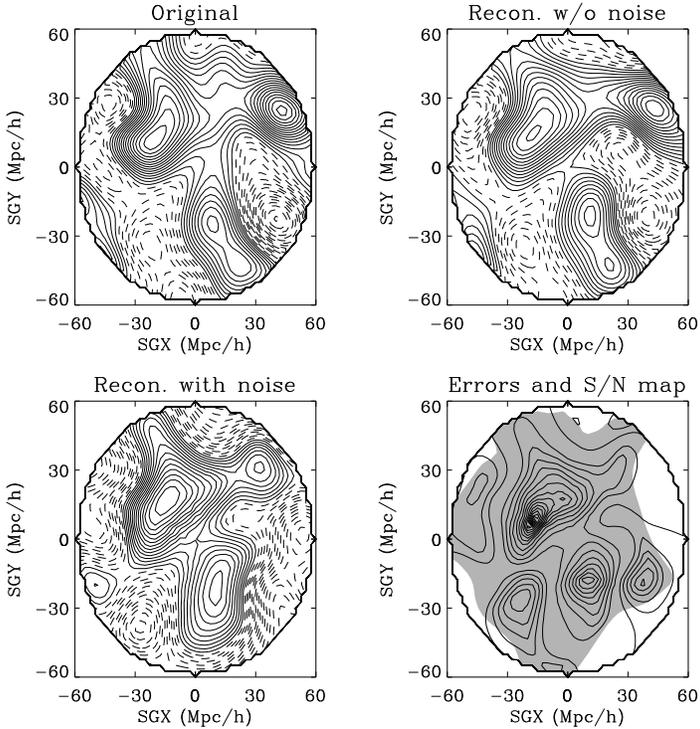}}
\end{picture}
\caption{Testing the method with mock SEcat data.  Shown are maps of
density in the Supergalactic plane, smoothed with a Gaussian window of
$900\ {\rm km s^{-1}}$, G9. Density contour spacing is $0.1$, the mean
$\delta=0$ contour is heavy, positive contours are solid and negative
contours are dashed.  Top-left panel shows the original mock
Supergalactic-plane. Top-right panel is the reconstruction from a
noiseless mock catalog, which shows the uniformity of the sampling and
the quality of the interpolation.  Bottom-left panel shows a typical
UMV reconstruction from noisy data.  Bottom-right panel shows the
signal--to--noise ratio with contour spacing 1., the heavy-solid line
indicate signal--to--noise ratio of unity. The shading indicates
regions where the error is less than 0.45}
\label{fig:test}
\end{figure}

We wish to test two aspects of the reconstruction. First, whether the
coverage within the assigned area is good enough for a faithful
recovery of the underlying signal. Second, whether the noise level
allows a reasonable (high Signal/Noise) reconstruction within the $60
\hmpc$ sphere. The success of the reconstruction is demonstrated in
Figure~\ref{fig:test}. The top-left panel shows the density of the
underlying field, used to construct the mock SEcat catalog, at the
$Z=0 \hmpc$ plane, smoothed with $9 \hmpc$ Gaussian, within a sphere
of radius $60 \hmpc$; this is the target map which we are attempting
to recover. 

In order to test the quality of the coverage we construct a noise-free
mock SEcat catalog which has accurate radial velocities at the
locations of the data points. The top-right panel in
Figure~\ref{fig:test} shows the density reconstruction from the
noise-free catalog, this map tests mainly the uniformity of the
catalog within a sphere of $60 \hmpc$. The very good agreement between
this map and the target map shows that the coverage of the SEcat
catalog is excellent.  

\begin{figure}
\setlength{\unitlength}{1cm} \centering
\begin{picture}(8,4)
\put(-1.5, -1.5){\includegraphics{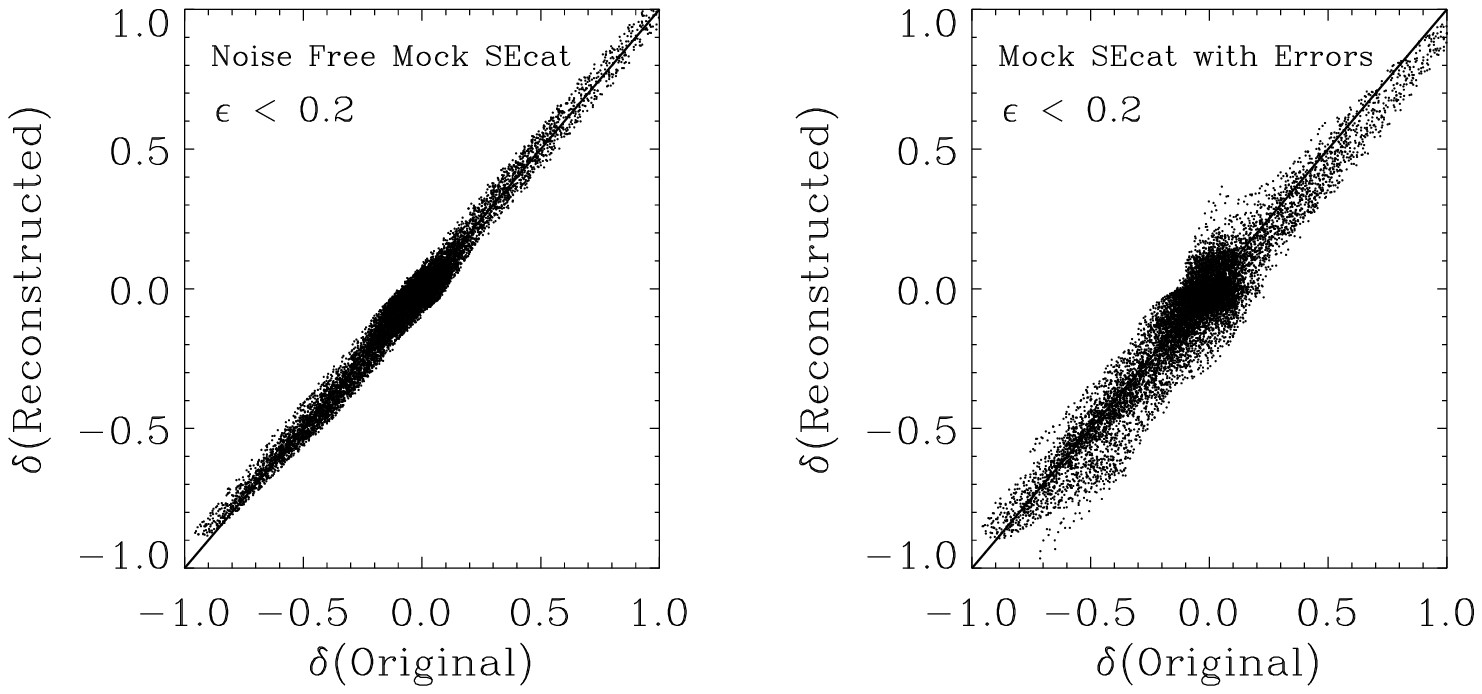}}
\end{picture}
\caption{Shown is scatter plot of the underlying density {\it vs.} the
reconstructed density. The densities chosen are from areas within
which the reconstruction error is $< 0.2$, this error is determined
using  30 mock SEcat catalogs. Left panel shows the quality of the
reconstruction from a noise-free mock catalog. Right panel shows the quality of the
reconstruction from a noisy mock SEcat catalog.}
\label{fig:den_scatter}
\end{figure}

Next, we test the effect of noise on the reconstruction; here we
construct a noisy mock SEcat catalog where the added noise is
realistic and corresponds to the quoted Tully-Fisher and $D_n-\sigma$
errors in the real catalog. The lower-left panel shows density
reconstruction from a noisy mock realization of the SEcat catalog.
The agreement between this map and the target map is also good (see
Figure~\ref{fig:den_scatter}), though there are some areas where the
recovered signal is not very satisfying, especially towards the edge
of the sphere. 

In order to test where the recovered signal is reliable we reconstruct
the density field from 30 mock catalogs, with the same underlying
field but with different noise realizations, and compare them with the
target map. The lower-right panel shows contours of the
signal--to--noise ratio, with spacing of 1 and with heavy-solid
contour denotes a signal--to--noise ratio of unity. The
signal--to--noise ratio in some areas in the map can get up to 15 and
in most of the map the signal--to--noise ration is greater than a
few. However, if the underlying density is of order zero the
signal--to--noise ratio gives a misleading impression about the
quality of the reconstruction as in this case it will be always of the
order of zero. Therefore, the lower-right panel also shows the area
(shaded) within which the error in the density-contrast is less than
0.45.

\begin{figure}
\setlength{\unitlength}{1cm} \centering
\begin{picture}(8,14)
\put(-2, -2.5){\includegraphics{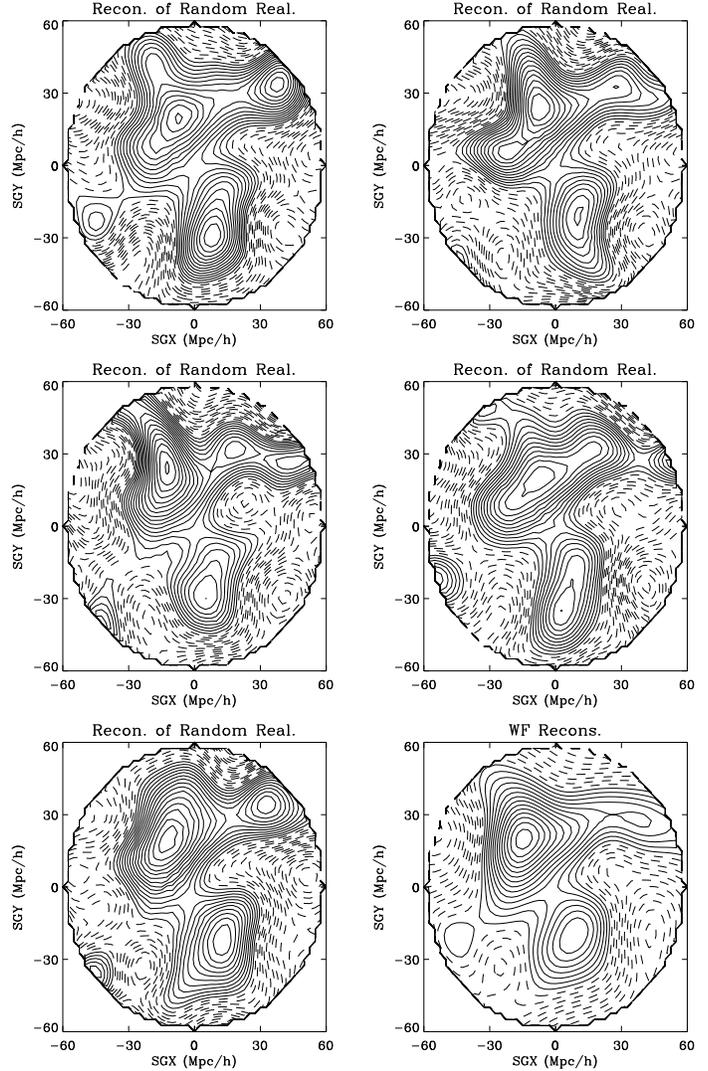}}
\end{picture}
\caption{Shown are maps of density in the Supergalactic plane
reconstructed from different Monte-Carlo realizations of the errors
The smoothing window is a Gaussian of radius $900\ {\rm km s^{-1}}$,
G9. Density contour spacing is $0.1$, the mean $\delta=0$ contour is
heavy, positive contours are solid and negative contours are
dashed. The bottom-right panel shows the Supergalactic plane WF
reconstruction from one of the Monte-Carlo realizations.}
\label{fig:random}
\end{figure}

To demonstrate the stability of the inversion, all the panels, with
the exception of the lower-right panel, in figure~\ref{fig:random}
show reconstructions similar to the ones shown in the lower-left panel
of fig~\ref{fig:test} but with different error realizations. For
comparison, the lower-right panel in fig.~ \ref{fig:random} shows the
Wiener reconstructed density field from one of the noisy mock catalogs
(the one used on the top left panel) , note that here the WF
reconstruction roughly recovers the features in the target map;
however, the amplitude of the density is suppressed throughout the
plane, reflecting the biased nature of the Wiener reconstruction.

Figure~\ref{fig:den_scatter} shows a scatter plot of the original {\it
vs.} reconstructed densities within the whole reconstructed
sphere. The densities are chosen from areas within which the errors,
estimated from Monte Carlos of noisy mock SEcat catalogs, are less
than 0.2. The left panel shows the quality of the reconstruction from
noise free catalog. While the right panel shows the reconstruction
quality from noisy catalog. As expected, the scatter in the right
panel is larger. The measured slope is slightly smaller than 1
($\approx 0.98 \pm 0.03 $, the quoted uncertainty is the $1\sigma$
error), nevertheless the agreement between the original and
reconstructed is excellent.

\section {Discussion \& Summary}
\label{sec:summary}

A general framework of linear estimation and prediction by minimal
variance subject to a linear constraint in the data has been
introduced. The solution of the minimization problem yields the UMV
estimator, which is shown to be a very useful tool for reconstructing
the large scale structure of the universe from incomplete, noisy and
sparse data.  The UMV estimator has been designed to overcome one of
the main drawbacks of the WF which is that it predicts the null field
in the absence of good data, \ie\ in the limit of very poor
signal-to-noise data the cosmological mean field is estimated, \ie\
zero perturbation. In contrast, the UMV estimator does not alter the
values of the reconstructed field at the locations of the data point,
because it lacks the filtering aspect of the WF, instead it keeps the
values of the measured data at the locations of the data points and
interpolates between them in accordance with the correlation function
assumed in the model. Like the WF the new estimator
can be used for dynamical reconstruction, \ie\ to reconstruct one
dynamical field, \eg\ over-density, from another measured field,
\eg\ radial peculiar velocity. These two properties make the UMV
estimator a very appealing tool for various applications in LSS and
CMB problems.

The regularization strength of the new estimator stems from the cross
correlation between the signal and the data points at different
locations. Which allows a reconstruction that is consistent with the
data points and the correlations among them. Lacking such a
correlation, the UMV estimator is equivalent to direct inversion. In
this case an additional regularization is required.  This issue could
be viewed as follows. The UMV produces an unbiased estimate of the
underlying signal subject to the minimal variance requirement. This
minimal variance could sometimes be too large to be useful. In
comparison the WF produces an {\it truly} minimal variance but with a
biased mean. Therefore, when applying the UMV estimator, sometimes an
extra regularzation is needed in order to, on the one hand, reduce the
variance in the UMV reconstruction to manageable value, while keeping
the reconstructed underlying signal as unbiased as possible, on the
other. In this paper two extra regularization methods, that are both
simple and easily integrated in the UMV method, have been discussed.

Constrained realizations of the underlying field (Hoffman \& Ribak
1991) will not be possible with the UMV estimator. However,
constrained realizations of a noisy underlying field are
possible. Such noisy constrained realizations could be very useful if
one wishes to fill gaps in the data, \eg\ from balloon-born CMB
measurements, with a realization that has the same assumed properties
of the Signal+Noise.

An apparent difficulty arises from the fact that the current UMV
reconstruction assumes linear gravitational instability, yet it is
applied to a universe that is {\it non-linear} on scales smaller than
a few Mpc's.  To obtain non-linear reconstruction of the
underlying field, $\bf s_{nl}$, one can include the non-linear
correlation by substituting $\langle {\bf s_{nl} o^+} \rangle$ in
eq.~\ref{eq:estimator}. This of course would only take into account
the contribution of the variance and ignore the contribution of higher
moments but in many cases this will do.

In one of the exampled presents here, we have applied the method
to galaxy radial peculiar velocity data and shown that the
reconstruction is unbiased and trustworthy within a very large region
of the volume covered by the data. In this context the UMV
reconstruction could be viewed as a compromise between the POTENT
algorithm (Bertschinger \& Dekel 1989; Dekel, Bertschinger \& Faber
1990), which assumes no regularization, and the WF, which relies too
heavily on it.

The UMV reconstruction of the over-density and the 3D velocity field
from galaxy peculiar velocity catalogs is suitable for bias parameter
extraction from a comparison with the respective fields obtained from
redshift galaxy catalogs like the PSCz (Saunders \etal\ 2000; Branchini
\etal\ 2000). The current estimator allows carrying out density-density
and velocity-velocity comparisons using the same reconstruction
technique (Zaroubi \etal\ 2001b).

\section{Acknowledgements}

I would like to thank A.J. Banday, E. Branchini,Y . Hoffman, \& H. Mo
for detailed comments on the manuscript and L.N. da Costa, O. Lahav,
T. Theuns, and A. Nusser for discussions. I also thank the anonymous
referee whose excellent comments have led to significant improvement
of the paper.

{}

\end{document}